\documentclass[sigconf]{acmart}
\AtBeginDocument{%
  }

\copyrightyear{2026}
\acmYear{2026}
\setcopyright{cc}
\setcctype{by}
\acmConference[WACE '26]{WebAssembly, Cloud and Edge: Shaping the Future of Computing}{December 14--18, 2026}{Tarragona, Spain}
\acmBooktitle{WebAssembly, Cloud and Edge: Shaping the Future of Computing (WACE '26), December 14--18, 2026, Tarragona, Spain}
\acmDOI{10.1145/3861044.3862582}
\acmISBN{979-8-4007-3109-9/2026/12}

\begin{document}

%%
%% The "title" command has an optional parameter,
%% allowing the author to define a "short title" to be used in page headers.
\title{Reducing Cold-Start Latency on the WebAssembly Edge via Locality and Network-Aware Scheduling}

\author{Michiel Van Kenhove}
\orcid{0000-0003-3623-2754}
\correspondingauthor
\affiliation{%
    \institution{IDLab, Ghent University - imec}
    \city{Ghent}
    \country{Belgium}
}
\email{michiel.vankenhove@ugent.be}

\author{Emiel Sucaet}
\orcid{0009-0004-8887-4630}
\author{José Santos}
\orcid{0000-0002-6276-2057}
\affiliation{%
    \institution{IDLab, Ghent University - imec}
    \city{Ghent}
    \country{Belgium}
}

\author{Filip De Turck}
\orcid{0000-0003-4824-1199}
\author{Bruno Volckaert}
\orcid{0000-0003-0575-5894}
\author{Merlijn Sebrechts}
\orcid{0000-0002-4093-7338}
\affiliation{%
    \institution{IDLab, Ghent University - imec}
    \city{Ghent}
    \country{Belgium}
}

%%
%% By default, the full list of authors will be used in the page
%% headers. Often, this list is too long, and will overlap
%% other information printed in the page headers. This command allows
%% the author to define a more concise list
%% of authors' names for this purpose.
\renewcommand{\shortauthors}{Van Kenhove et al.}

%%
%% The abstract is a short summary of the work to be presented in the
%% article.
\begin{abstract}
Edge computing places computation closer to users and data sources, but introduces heterogeneous hardware, limited resources, and variable network conditions. WebAssembly (Wasm) is increasingly considered for these environments because modules are compact, portable, and quick to start. However, cold-start latency remains a challenge: when a worker node does not yet hold the required artifact, it must fetch it before execution, adding transfer overhead that directly increases the end-to-end (E2E) latency, especially for large artifacts or slow network links. At the same time, the geographical distance between the requester and the assigned worker independently contributes to latency through network routing overhead. Together, artifact locality and network proximity do not always point to the same placement decision, and naive scheduling strategies fail to account for either. This work extends a Wasm-native edge orchestrator with state-aware scheduling strategies that jointly optimize for both objectives. Evaluation on a Raspberry Pi testbed with emulated regional network latency shows that a hybrid scheduler combining cache-state and network-proximity signals reduces mean E2E latency by up to 40.8\% compared to a default round-robin baseline. Artifact locality-aware placement is most effective when artifacts are large or repeatedly requested, while network-aware placement dominates when artifacts are small and routing latency has a larger relative effect, but combining both objectives consistently yields the lowest E2E latency across diverse edge workloads.
\end{abstract}

%%
%% The code below is generated by the tool at http://dl.acm.org/ccs.cfm.
%% Please copy and paste the code instead of the example below.
%%


%%
%% Keywords. The author(s) should pick words that accurately describe
%% the work being presented. Separate the keywords with commas.
%% A "teaser" image appears between the author and affiliation
%% information and the body of the document, and typically spans the
%% page.
% \begin{teaserfigure}
%  \includegraphics[width=\textwidth]{sampleteaser}
%   \caption{Seattle Mariners at Spring Training, 2010.}
%   \Description{Enjoying the baseball game from the third-base
%   seats. Ichiro Suzuki preparing to bat.}
%   \label{fig:teaser}
% \end{teaserfigure}

% \received{20 February 2007}
% \received[revised]{12 March 2009}
% \received[accepted]{5 June 2009}

%%
%% This command processes the author and affiliation and title
%% information and builds the first part of the formatted document.
\maketitle

\section{Introduction}

Modern applications increasingly shift computation away from centralized cloud data centers and toward the network edge, placing execution closer to users, sensors, and other data sources across distributed environments~\cite{satyanarayananEmergenceEdgeComputing2017}. This transition to the edge-cloud continuum is driven by the rapid growth of Internet of Things (IoT) applications, real-time analytics, and latency-sensitive services~\cite{gkonisSurveyIoTEdgeCloudContinuum2023,belcastroNavigatingEdgeCloudContinuum2026}. While edge computing effectively reduces the physical distance between data generation and execution, it introduces complex scheduling challenges~\cite{luoResourceSchedulingEdge2021,soumplisResourceAllocationChallenges2022}. This complexity arises primarily because individual edge nodes are often resource-constrained and operate under less controlled conditions than centrally managed infrastructure, with limited compute and memory capacity, variable network connectivity, and no guaranteed physical redundancy. Consequently, traditional heavyweight container deployments are often unsuitable for these environments~\cite{yakubovComparativeAnalysisLightweight2025}, especially for latency-sensitive applications where a primary scheduling objective is to minimize end-to-end (E2E) task latency.

WebAssembly (Wasm) is increasingly considered as a lightweight alternative for this setting~\cite{menetreyWebAssemblyCommonLayer2022}. Wasm modules are compact, por\-ta\-ble binary artifacts that can be executed by compatible runtimes across heterogeneous platforms without recompilation~\cite{haasBringingWebSpeed2017a}. Compared to container-based deployments, Wasm can reduce cold-start latency, memory overhead, and artifact size across a range of serverless and edge scenarios~\cite{gackstatterPushingServerlessEdge2022,kjorveziroskiWebAssemblyOrchestrationContext2023}. These properties reduce several sources of deployment overhead, making Wasm a promising solution for latency-sensitive edge workloads. However, achieving low E2E latency still depends on how workloads are scheduled across the edge.

This work builds on Propeller~\cite{abstractmachinesPropellerOpenSourceWebAssembly}, an open-source~\cite{abstractmachinesAbsmachPropeller2026} Wasm-native orchestrator for the edge-cloud continuum. Unlike Ku\-ber\-net\-es-integrated~\cite{kubernetesauthorsKubernetesProductionGradeContainer} approaches such as wasmCloud~\cite{WasmCloudWebAssemblyApplication} and Spin\-Kube~\cite{SpinKube}, which extend existing container-oriented infrastructure to support Wasm workloads, Propeller is designed from the ground up around edge Wasm workloads. At a high level, Propeller consists of a central Manager that coordinates task scheduling and distributed worker nodes called Proplets that execute Wasm workloads. The baseline Propeller scheduler uses round-robin assignment.

Propeller's naive round-robin scheduler leaves two independent sources of E2E task latency unaddressed. When a Proplet has not previously executed a given task, it must first fetch the required artifact from the registry before execution can begin, a cold-start overhead that compounds with artifact size and request frequency~\cite{joosenServerlessColdStarts2025,golecColdStartLatency2024}. Independently, the round-robin scheduler does not consider the network latency between the requester and the assigned Proplet, which may add significant network routing overhead on both the request and result paths. Artifact locality and network proximity do not always favor the same placement decision, and stateless scheduling strategies such as round-robin ignore both.

To address this, this work extends Propeller with local artifact caching and two state-aware scheduling strategies that jointly address both latency sources, combined in a hybrid scheduler. Specifically, the contributions in this paper are:

\begin{enumerate}
    \item a per-Proplet \textbf{local artifact cache} with cache-state reporting to the Manager, eliminating redundant artifact transfers and providing the foundation for locality-aware placement decisions;
    \item a \textbf{locality-aware} scheduler that tracks which artifacts are already cached on each Proplet and prioritizes warm workers to avoid redundant artifact transfers;
    \item a \textbf{network-aware} scheduler that selects Proplets based on estimated worker-to-requester latency, with direct result routing back to the requester;
    \item a \textbf{hybrid} scheduler that jointly optimizes both objectives through a cost model combining a download penalty and a routing latency penalty, and proactively prefetches the artifact to the lowest-latency cold Proplet to warm it for future requests.
\end{enumerate}

The remainder of this paper is structured as follows. Section~\ref{sec:design-implementation} presents the proposed state-aware scheduling strategies and their implementation in Propeller, while Section~\ref{sec:evaluation} evaluates the schedulers across different workload scenarios. Section~\ref{sec:related-work} situates these results within related work. Finally, Section~\ref{sec:conclusion} concludes the paper.

\section{State-Aware Scheduling}\label{sec:design-implementation}

Propeller consists of a central Manager, Proplets, a Proxy, and an MQTT message broker~\cite{propeller_docs_system_architecture}. The Manager handles task management, workflow orchestration, and Proplet coordination via a REST API, assigning tasks to available Proplets. Proplets execute Wasm workloads, while the Proxy fetches artifacts from an OCI registry and forwards them over MQTT.

Task placement in the baseline architecture uses round-robin scheduling, which provides simple load distribution but ignores both artifact locality and worker-to-requester network proximity. Proplets also lack a local artifact cache, causing redundant transfers on repeated invocations. These limitations motivate the state-aware extensions introduced below: local artifact caching, artifact locality-aware scheduling, network-aware scheduling, and hybrid scheduling with prefetching.

\subsection{Local Artifact Caching}

To reduce redundant artifact transfers, a local Wasm artifact cache was added to each Proplet. Whenever a task starts, the assigned Proplet first checks whether the required artifact is available locally. On a cache hit, the workload is loaded directly from disk. On a cache miss, the Proplet requests the artifact from the Proxy and stores it for later reuse. This cache is a generic runtime enhancement, so cache hits may occur under any scheduler, including round-robin and network-aware scheduling, depending on prior task placements and execution history.% Artifact delivery was also made Proplet-specific rather than broadcast over a shared MQTT topic, avoiding redundant chunk transfers to nodes that did not request the artifact.

\subsection{Locality-Aware Scheduling}

While the local cache can reduce artifact-transfer overhead under any scheduler, the locality-aware scheduler explicitly incorporates cache state into placement decisions. Using the cache-state updates reported by Proplets, the Manager classifies candidate Proplets as warm if they already cache the requested artifact, or cold otherwise. If at least one warm Proplet is available, the scheduler selects the least-loaded warm candidate. If no warm candidate exists, it falls back to the least-loaded cold candidate. This policy preserves simple load balancing while prioritizing artifact reuse.

\subsection{Network-Aware Scheduling}\label{sec:network-aware-scheduling}

Artifact locality alone does not minimize E2E latency in geographically distributed edge systems. A warm Proplet may be far from the requester, while a nearby Proplet may require a cold start. To account for worker-to-requester network proximity, task submissions include a requester region and each Proplet is assigned a region. Candidate Proplets are scored using a static latency matrix derived from inter-region latency measurements~\cite{AWSLatencyMonitoring}, and the scheduler selects the worker with the lowest estimated Proplet-to-requester latency, applying load-based tie-breaking when multiple candidates have the same latency.

This placement decision is only effective if the result path also avoids unnecessary detours. In the original architecture, results are routed back through the Manager regardless of where the task executed, creating a central bottleneck even when the Proplet is close to the requester. To preserve the latency benefit of network-proximity-aware placement, tasks may include a callback URL, allowing the selected Proplet to return the user-facing result directly to the requester, while still reporting completion to the Manager.

\subsection{Hybrid Scheduling and Prefetching}

\begin{figure}[t]
    \centering
    \includegraphics[width=\columnwidth]{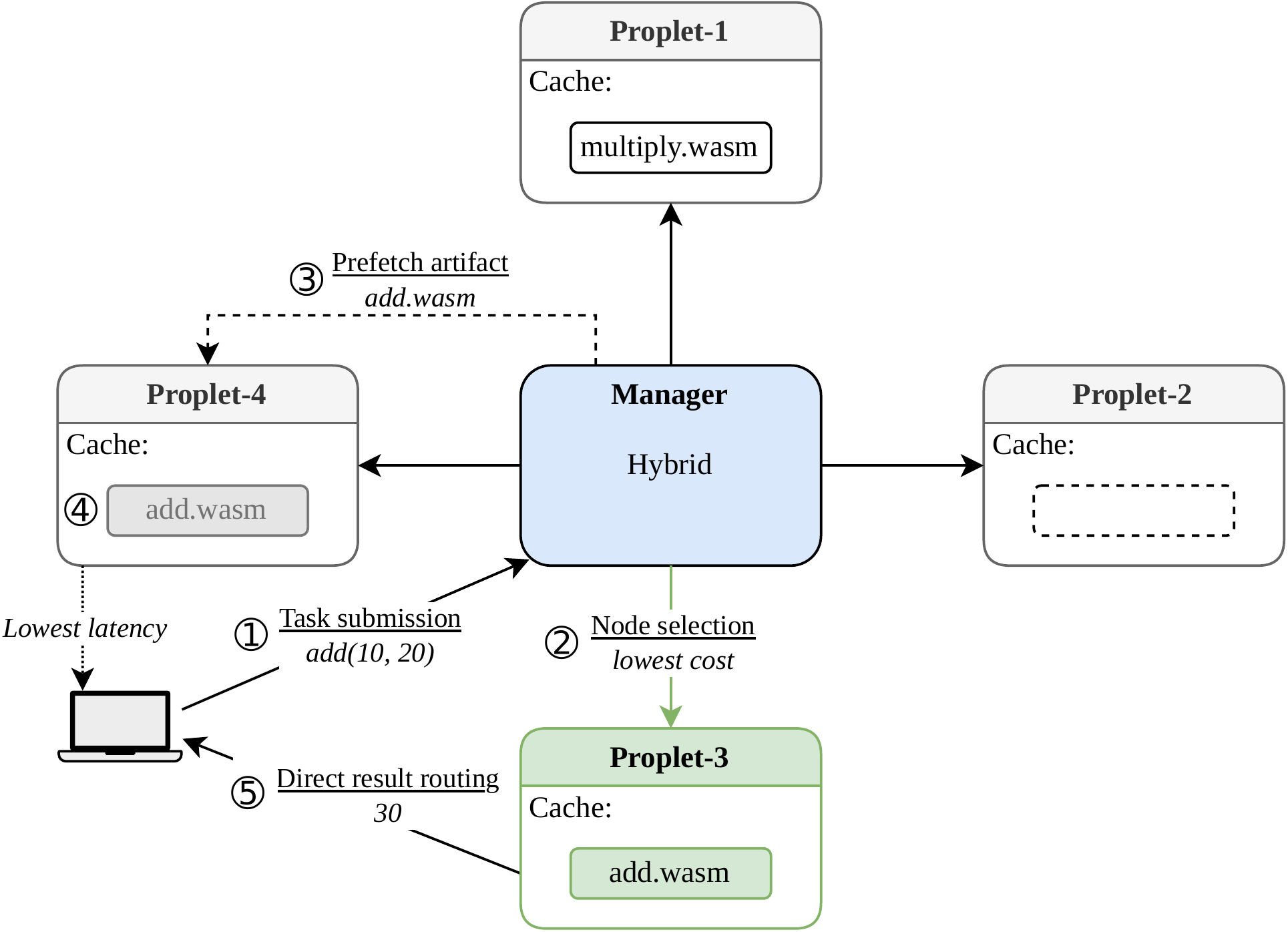}
    \caption{Overview of the proposed hybrid scheduler in action. Task submission~\ding{172} triggers cost-based worker node selection, where Proplet-3 wins on total cost due to its warm cache~\ding{173}. In parallel, the Manager prefetches~\ding{174} the artifact to Proplet-4, the lowest-latency node for this requester, warming it for future requests~\ding{175}, while Proplet-3 routes the result directly back to the user~\ding{176}.}\label{fig:hybrid-scheduler}
    \Description[Diagram showing the Manager selecting the lowest-cost Proplet for task execution while prefetching the artifact to the lowest-latency Proplet, with the result routed directly back to the requester.]{A block diagram with a central blue box labeled "Manager, Hybrid" connected to four Proplet worker boxes and a laptop icon representing the requester. Proplet-1 (top) shows a cache containing multiply.wasm. Proplet-2 (right) shows an empty cache. Proplet-4 (left) shows an empty cache, later filled with add.wasm. Proplet-3 (bottom, outlined in green) shows a cache containing add.wasm. A laptop icon at the bottom left sends a task submission, add(10, 20), to the Manager, labeled step 1. A green arrow from the Manager to Proplet-3 is labeled step 2, node selection, lowest cost. A dashed arrow from the Manager to Proplet-4 is labeled step 3, prefetch artifact, add.wasm, and Proplet-4's cache is marked step 4 once add.wasm arrives, with a dotted arrow labeling Proplet-4 as the lowest-latency node. A final arrow from Proplet-3 back to the laptop is labeled step 5, direct result routing, 30, bypassing the Manager.}
\end{figure}

The locality-aware and network-aware schedulers optimize different objectives and can therefore make conflicting placement decisions. The hybrid scheduler combines both signals through a simple additive cost model, as illustrated in Figure~\ref{fig:hybrid-scheduler}. A task submission for \texttt{add(10, 20)} arrives at the Manager~\ding{172}. The Manager scores each candidate Proplet using the combined cost model and selects the node with the lowest total cost~\ding{173}, in this case Proplet-3, which already has a cached version of \texttt{add.wasm}, avoiding a cold-start transfer entirely. Because the Proplet with the lowest latency to the requester (Proplet-4) is cold, the Manager concurrently issues a prefetch message~\ding{174}, so a future request from the same region can be served with both a warm cache and low routing latency~\ding{175}. Once execution completes, Proplet-3 routes the result directly to the user~\ding{176}, bypassing the Manager and avoiding an extra network hop on the return path.

For a task $T$ and candidate Proplet $P$, its estimated cost is

\begin{equation}
C(T,P) = C_{\mathrm{download}}(T,P) + C_{\mathrm{routing}}(T,P).
\end{equation}

The download component is zero when the requested artifact is already cached on $P$. Otherwise, it is assigned a heuristic cold-start penalty based on the artifact class. These penalties were derived from a calibration experiment on the same testbed used for the evaluation in Section~\ref{sec:evaluation}. For each artifact class, 100 cold and 100 warm round-robin executions were measured, yielding a mean cold-to-warm latency difference of approximately 150 ms for small artifacts and 677 ms for large artifacts. Accordingly, we used a penalty of 150 ms for small artifacts and rounded the large-artifact penalty to 700 ms. These two classes are sufficient to capture the dominant cost difference observed in our evaluation, and we leave a continuous, throughput-based cost estimator to future work. The routing component is the estimated latency between the requester region and the Proplet region. The hybrid scheduler selects the candidate with the lowest total cost.

This model allows the scheduler to adapt to workload characteristics. For small artifacts, the lower download penalty means that a nearby cold Proplet may still be preferred. For large artifacts, avoiding a transfer can dominate, so a warm but more distant Proplet may be selected. To improve future decisions, the hybrid scheduler also performs asynchronous prefetching. If the hybrid scheduler selects a Proplet other than the lowest-latency candidate, the Manager asynchronously prefetches the artifact to that candidate if it is not already cached there. As a result, later requests from the same region benefit from both artifact locality and network proximity.

\section{Evaluation Results}\label{sec:evaluation}

The proposed schedulers were evaluated on a small-scale edge testbed consisting of a centralized control plane and four Proplet worker nodes deployed on four Raspberry Pi 4B devices with 2 GB of RAM. The control plane hosted the Propeller Manager, Proxy, MQTT broker, local OCI registry, and benchmarking tools. All nodes were connected through a Gigabit Ethernet network.

To study geographically distributed deployments under repeatable conditions, regional network delays were emulated using Linux traffic control with \texttt{tc}~\cite{Tc8LinuxManual} and \texttt{netem}~\cite{Tcnetem8LinuxManual}. The emulation separated cloud-to-edge delay from edge-to-user delay, making it possible to model both artifact-transfer overhead and worker-to-requester routing latency. Each Proplet worker was assigned a fixed region, while requester regions were specified per task. The corresponding worker-to-requester latency was estimated from a regional latency matrix based on AWS inter-region measurements~\cite{AWSLatencyMonitoring}.

The workloads submitted to the testbed draw from two artifact classes: small Wasm artifacts of approximately 100 KiB and large artifacts of approximately 5 MiB, with large artifacts representing network-bound workloads where transfer cost dominates. The experiments varied the task-universe size, requester-location distribution, large-artifact ratio, and sequence length. The task-universe size determined how many distinct Wasm workloads could recur within a sequence. Requester locations were drawn from one of three scenarios: \textit{local}, where requests originate from regions close to one of the available Proplet workers; \textit{remote}, where requests originate from regions without a directly corresponding worker; and \textit{mixed}, which combines both request origins within the same sequence. Each experiment consisted of 10 repetitions, where for each repetition one pseudo-random task sequence was generated and executed once with each scheduler. Unless otherwise stated, each benchmark run consisted of a sequence of 40 task submissions drawn from a task universe of 4 under the mixed requester-location scenario. Proplet caches were cleared before each scheduler run to enforce a cold-cache start. E2E latency is reported as the primary metric, measured from task submission until the result is received at the region-specific callback endpoint. All four schedulers used the same direct result-routing mechanism described in Section~\ref{sec:network-aware-scheduling}, so differences in reported latency reflect placement decisions rather than differences in result routing.

\subsection{Effect of Large-Artifact Ratio}

\begin{figure*}[t]
    \centering
    \includegraphics[width=1.0\textwidth]{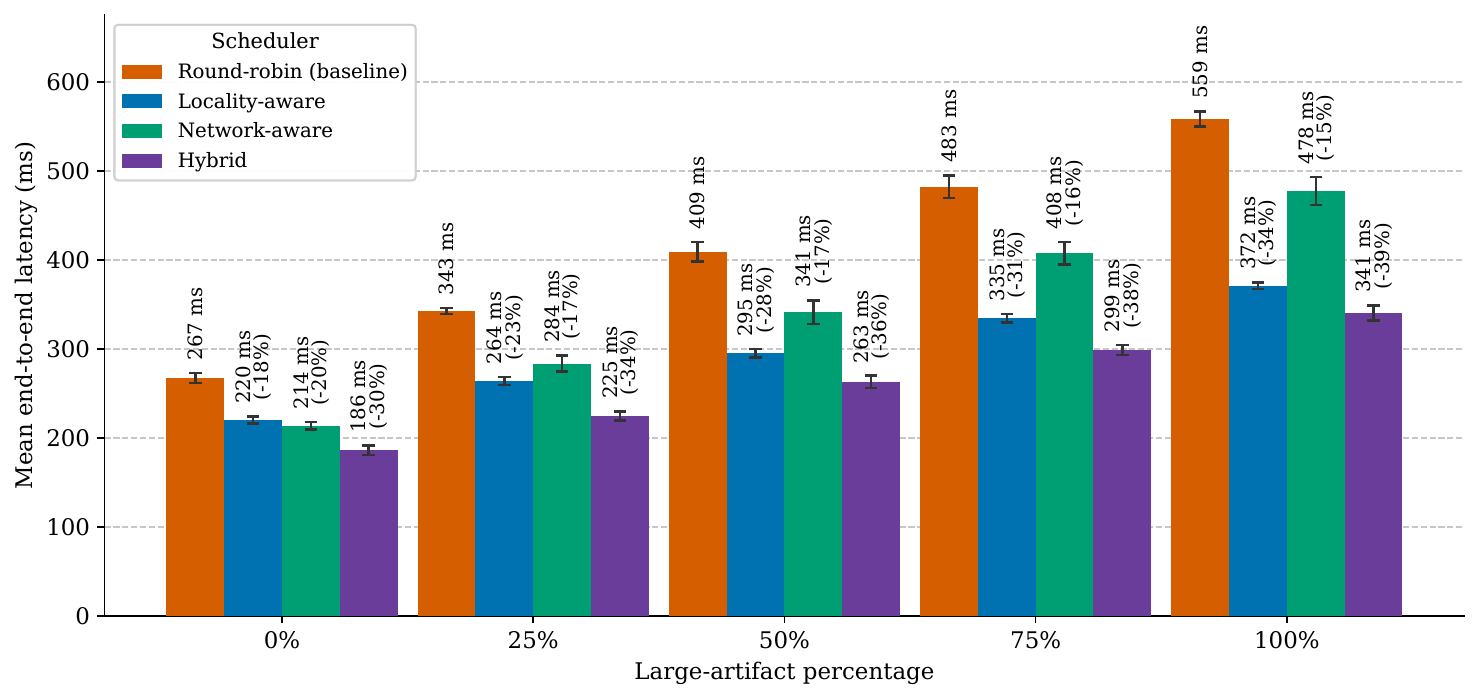}
    \caption{Mean E2E latency per submitted task as the large-artifact share of the sequence increases under the mixed requester-location scenario. Error bars indicate the 95\% confidence interval. As artifact-transfer costs grow, locality-aware scheduling increasingly improves over round-robin, whereas the hybrid scheduler consistently achieves the lowest latency by balancing artifact locality with requester network proximity.}\label{fig:large-artifact-percentage}
    \Description[Grouped bar chart showing mean end-to-end latency (ms) for four schedulers: round-robin, locality-aware, network-aware, and hybrid; across five large-artifact percentages (0\%, 25\%, 50\%, 75\%, 100\%). Latency rises for all schedulers as large-artifact share increases, but the hybrid scheduler stays lowest throughout, cutting latency 30-39\% versus round-robin.]{Grouped bar chart with the x-axis showing large-artifact percentage in the task sequence (0\%, 25\%, 50\%, 75\%, 100\%) and the y-axis showing mean end-to-end latency in milliseconds, ranging from 0 to 600 ms, with error bars indicating 95\% confidence intervals. Each x-axis group contains four bars representing the round-robin baseline, locality-aware, network-aware, and hybrid schedulers. Round-robin latency increases most sharply as the large-artifact share grows, from 267 ms at 0\% to 559 ms at 100\%. Locality-aware scheduling starts at 220 ms (18\% below round-robin) and rises more gently to 372 ms at 100\% (34\% below round-robin), reflecting its growing advantage as artifact-transfer costs increase. Network-aware scheduling shows the opposite pattern: it performs strongly at 0\% large artifacts (214 ms, 20\% below round-robin) but its relative advantage shrinks as large-artifact share increases, reaching 478 ms at 100\% (only 15\% below round-robin). The hybrid scheduler achieves the lowest latency at every percentage, ranging from 186 ms at 0\% (30\% below round-robin) to 341 ms at 100\% (39\% below round-robin), consistently outperforming all other schedulers across the full range.}
\end{figure*}

Figure~\ref{fig:large-artifact-percentage} evaluates how each scheduler behaves as artifact-transfer costs become increasingly important. The experiment varies the share of large artifacts in the task sequence from entirely small artifacts to entirely large artifacts under the mixed requester-location scenario. As the large-artifact ratio increases, cache misses become more expensive, stressing schedulers that do not account for artifact locality.

Mean E2E latency increases as the sequence shifts from small to large artifacts, with round-robin affected most strongly. Its latency more than doubles, from 267~ms for all-small sequences to 559~ms for all-large sequences, because it ignores cache state and often assigns tasks to Proplets that must fetch the requested artifact before execution. Locality-aware scheduling reduces this overhead by prioritizing warm Proplets, and its advantage over round-robin grows as artifact transfers become more expensive. The improvement increases from 17.6\% for all-small sequences to 33.5\% for all-large sequences, confirming that cache-aware placement becomes more valuable when artifact-transfer overhead grows.

The network-aware scheduler shows the opposite trend. It performs well when artifacts are small, reaching 214~ms, but its benefit decreases as large artifacts become more frequent. With only large artifacts, mean latency rises to 478~ms, showing that requester network proximity alone cannot compensate for repeated cold artifact transfers.

The hybrid scheduler achieves the lowest latency for every large-artifact percentage, increasing from 186~ms to 341~ms across the same range. This corresponds to a 30.2\%--39.0\% reduction compared to round-robin. Overall, the results show that the hybrid cost model adapts to sequence composition by favoring nearby workers when transfers are cheap and warm caches when artifact downloads become expensive.

\subsection{Effect of Requester Locality}

\begin{figure}[t]
    \centering
    \includegraphics[width=1.0\columnwidth]{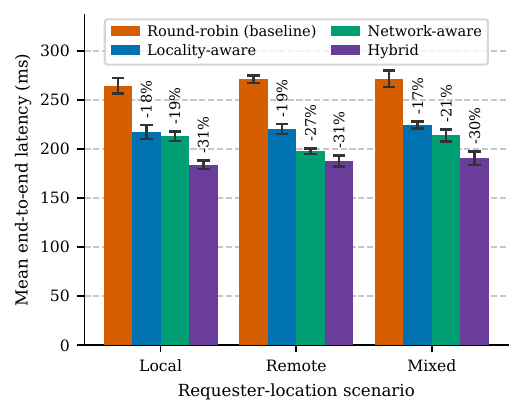}
    \caption{Mean E2E latency per task across local, remote, and mixed requester-location scenarios for small-artifact workloads. Error bars indicate the 95\% confidence interval. Because artifact-transfer costs are minimized in this experiment, the results primarily expose the impact of worker-to-requester network proximity. Network-aware scheduling consistently improves over round-robin by selecting lower-latency Proplets, while the hybrid scheduler achieves the lowest latency in all requester scenarios by combining network proximity with artifact locality.}\label{fig:requester-locality-latency}
    \Description[Grouped bar chart of mean end-to-end latency across Local, Remote, and Mixed requester scenarios, comparing four schedulers; Hybrid is lowest in all three.]{Grouped bar chart with three groups on the x-axis: Local, Remote, and Mixed requester-location scenarios; and mean end-to-end latency in milliseconds on the y-axis, ranging from 0 to 300 ms. Each group has four bars: Round-robin (baseline, orange), Locality-aware (blue), Network-aware (green), and Hybrid (purple), with 95\% confidence-interval error bars. In every scenario the bars decrease from Round-robin to Hybrid. In the Local scenario, Round-robin reaches 264 ms, dropping to about 216 ms for Locality-aware (-18\%), 213 ms for Network-aware (-19\%), and 184 ms for Hybrid (-31\%). The Remote scenario shows a similar pattern, with reductions of -19\%, -27\%, and -31\% relative to its Round-robin baseline, and a Hybrid latency of 188 ms. The Mixed scenario shows reductions of -17\%, -21\%, and -30\%, with a Hybrid latency of 190 ms. Across all three scenarios, Network-aware scheduling outperforms Locality-aware scheduling, since artifact-transfer costs are minimized (small artifacts only) and network proximity is the dominant latency factor, while Hybrid scheduling consistently achieves the lowest overall latency by combining both signals.}
\end{figure}

Figure~\ref{fig:requester-locality-latency} evaluates local, remote, and mixed requester-location scenarios using only small artifacts. By minimizing artifact-transfer costs, this setup isolates the effect of worker-to-requester network proximity. The mixed-scenario, all-small-artifact condition here matches the 0\% large-artifact condition in Figure~\ref{fig:large-artifact-percentage} in setup. The two figures report independent sets of repetitions, the small numeric differences between them are due to run-to-run variability.

The network-aware scheduler consistently reduces latency compared to round-robin, confirming that requester location significantly influences E2E response time. In the local scenario, round-robin reaches 264~ms, while network-aware scheduling reduces this to 213~ms. Similar improvements are observed in the remote and mixed scenarios, where selecting a lower-latency Proplet reduces the routing-overhead component. Locality-aware scheduling also improves over round-robin, although its advantage is smaller in this experiment because artifact downloads are relatively cheap. The hybrid scheduler again achieves the lowest latency in all three scenarios: 184~ms for local requesters, 188~ms for remote requesters, and 190~ms for the mixed scenario. These values correspond to reductions of approximately 30\% compared to round-robin. The result shows that even in workloads where network proximity is the dominant signal, retaining cache awareness remains beneficial.

\subsection{Additional Validation}

Further evaluations confirmed these trends under different workload conditions by varying the task-universe size and sequence length to assess how the schedulers behave under different cache-reuse conditions.

Varying the task-universe size $U$ changes the probability of cache reuse. With only large artifacts under a mixed requester-location scenario, round-robin latency increases from 442~ms for $U=2$ to 703~ms for $U=8$, because requests are more often assigned to Proplets without the required artifact. Locality-aware scheduling performs substantially better, reaching 321~ms, 368~ms, and 442~ms for $U=2$, $U=4$, and $U=8$, respectively. The hybrid scheduler remains best across all task-universe sizes, with mean latencies of 308~ms, 341~ms, and 416~ms, corresponding to improvements of 30.2\%, 39.2\%, and 40.8\% over round-robin. 

Varying the sequence length instead changes the amount of time it takes the system to warm the caches. With a mixed requester-location scenario and 50\% large artifacts, the hybrid scheduler decreases from 393~ms at 10 tasks to 278~ms at 30 tasks and 247~ms at 60 tasks. Compared to round-robin, this corresponds to reductions of 34.3\%, 38.9\%, and 32.3\%, respectively. The locality-aware scheduler also benefits from longer sequences, as cached artifacts are reused more often after the initial cold phase. The network-aware scheduler reduces routing overhead, but remains less effective than locality-aware and hybrid scheduling when large-artifact cache misses are still frequent.

Together, these additional evaluations show how cache-reuse conditions affect scheduler behavior. Larger task universes make cache misses more likely, while longer sequences create more opportunities for reuse after the initial cold phase. In both cases, the hybrid scheduler performs best by retaining the benefits of locality-aware placement while also accounting for requester network-proximity.

\section{Related Work}\label{sec:related-work}

Recent research has explored Wasm as a lightweight alternative to container-based deployments in edge and serverless computing. Compared to traditional containers, Wasm offers improved portability, reduced resource consumption, and lower startup overhead, making it attractive for heterogeneous edge environments. Gackstatter et al.~\cite{gackstatterPushingServerlessEdge2022} demonstrate that Wasm runtimes can significantly reduce cold-start latency and resource usage on constrained edge devices, consistent with our own findings at the orchestration level. More broadly, recent surveys and orchestration studies~\cite{kakatiWebAssemblyWebReview2023,kjorveziroskiWebAssemblyOrchestrationContext2023} highlight Wasm's potential for the edge-cloud continuum while identifying workload placement and orchestration as important open challenges, which this paper addresses directly.

Several studies seek to reduce startup overhead by exploiting artifact locality. S-Cache~\cite{chenSCacheFunctionCaching2023} incorporates function popularity, image size, and cold-start costs into both caching and request-distribution decisions, reducing redundant downloads and improving response latency. Our locality-aware scheduler follows a similar cache-state-driven philosophy and shows comparable gains once artifact size dominates the cost. Related work has similarly shown that artifact reuse and warm execution environments can substantially mitigate cold-start overheads in serverless platforms~\cite{shahradServerlessWildCharacterizing2020}. Other works instead focus on communication-aware workload placement. Diktyo~\cite{santosDiktyoNetworkAwareScheduling2023} incorporates network topology into Kubernetes scheduling decisions that reduce communication costs between distributed services, while GeoFaaS~\cite{malekabbasiGeoFaaSEdgetoCloudFaaS2024} routes invocations toward geographically appropriate edge nodes to reduce user-perceived latency. Our network-aware scheduler builds on the same principle and shows, consistent with these findings, its largest gains when artifacts were small and network latency dominated the response time.

Although these approaches demonstrate the benefits of artifact locality and network-aware scheduling, they typically optimize these objectives independently. Our results show that neither signal alone is sufficient across the workloads tested: the hybrid scheduler consistently ranked first across every configuration, confirming that artifact locality and network proximity should be treated as complementary rather than distinct objectives in Wasm-native edge orchestration.

\section{Conclusion}\label{sec:conclusion}

This paper set out to address a gap in Wasm-native edge orchestration: artifact locality and requester network proximity are both significant contributors to E2E task latency, yet naive schedulers such as round-robin account for neither, and prior work has largely treated the two objectives in isolation. By extending Propeller, a Wasm-native edge orchestrator, with a per-Proplet artifact cache, cache-state reporting, regional latency awareness, direct result routing, and asynchronous artifact prefetching, this work made both signals available to the scheduler and evaluated three strategies that exploit them individually and jointly.

The results confirm that neither signal alone is sufficient. Locality-aware scheduling scaled well with artifact-transfer cost, cutting latency by up to 33.5\% over round-robin as the share of large artifacts grew, but its advantage shrank to 17.6\% when artifacts were small and cheap to fetch. Network-aware scheduling showed the inverse pattern: it delivered the largest gains when artifacts were small and network routing latency dominated the response time, but its benefit diminished as cold artifact transfers became the primary source of latency. Across every workload configuration tested, varying artifact size, requester locality, task-universe size, and sequence length, the hybrid scheduler was the only strategy that consistently ranked first. It reduced mean E2E latency by approximately 30.2--40.8\% over round-robin by adapting its placement decision to whichever cost component was larger for a given task, and by proactively prefetching artifacts to warm colder-but-closer Proplets for future requests.

These findings indicate that artifact locality and network proximity should be treated as complementary rather than competing scheduling signals, and that a simple additive cost model is enough to capture their trade-off well in practice. This has a direct implication for Wasm edge orchestrators beyond Propeller: even lightweight, cache-unaware, or purely network-aware placement strategies leave measurable latency on the table, and combining both signals via a low-overhead cost function is a practical, generalizable way to close that gap.

The evaluation was conducted on a four-node Raspberry Pi testbed with emulated regional latency and synthetic Wasm artifacts, which made it possible to isolate and study the trade-off between artifact locality and network proximity under controlled, repeatable conditions. Building on this foundation, future work could further strengthen the case for state-aware scheduling in production settings by validating the hybrid scheduler on real multi-region infrastructure, incorporating bandwidth constraints and packet loss, moving from the current two-class download-cost heuristic to a continuous, throughput-based estimator, and introducing cache eviction policies for storage-constrained Proplets. These directions offer promising opportunities to extend state-aware scheduling toward broader, real-world Wasm edge deployments.

%%
%% The acknowledgments section is defined using the "acks" environment
%% (and NOT an unnumbered section). This ensures the proper
%% identification of the section in the article metadata, and the
%% consistent spelling of the heading.
\begin{acks}
This work received funding from the \grantsponsor{snsju-horizon}{Smart Networks and Services Joint Undertaking (SNS JU) under the Horizon Europe research and innovation program of the European Union}{https://doi.org/10.3030/101139067} under Grant Agreement number \grantnum[https://doi.org/10.3030/101139067]{snsju-horizon}{101139067} (\href{https://elasticproject.eu/}{https://elasticproject.eu/}). Expressed views and opinions are however those of the authors only and do not necessarily reflect those of the European Union. Neither the European Union nor the granting authority can be held responsible for them.

José Santos is funded by the \grantsponsor{jose-fwo}{Research Foundation Flanders (FWO)}{https://www.fwo.be/en/}, grant number \grantnum{jose-fwo}{1253226N}.
\end{acks}

%%
%% The next two lines define the bibliography style to be used, and
%% the bibliography file.
\bibliographystyle{ACM-Reference-Format}
\bibliography{references}

%%
%% If your work has an appendix, this is the place to put it.
% \appendix

% \section{Research Methods}

% \subsection{Part One}

% Lorem ipsum dolor sit amet, consectetur adipiscing elit. Morbi
% malesuada, quam in pulvinar varius, metus nunc fermentum urna, id
% sollicitudin purus odio sit amet enim. Aliquam ullamcorper eu ipsum
% vel mollis. Curabitur quis dictum nisl. Phasellus vel semper risus, et
% lacinia dolor. Integer ultricies commodo sem nec semper.

% \subsection{Part Two}

% Etiam commodo feugiat nisl pulvinar pellentesque. Etiam auctor sodales
% ligula, non varius nibh pulvinar semper. Suspendisse nec lectus non
% ipsum convallis congue hendrerit vitae sapien. Donec at laoreet
% eros. Vivamus non purus placerat, scelerisque diam eu, cursus
% ante. Etiam aliquam tortor auctor efficitur mattis.

% \section{Online Resources}

% Nam id fermentum dui. Suspendisse sagittis tortor a nulla mollis, in
% pulvinar ex pretium. Sed interdum orci quis metus euismod, et sagittis
% enim maximus. Vestibulum gravida massa ut felis suscipit
% congue. Quisque mattis elit a risus ultrices commodo venenatis eget
% dui. Etiam sagittis eleifend elementum.

% Nam interdum magna at lectus dignissim, ac dignissim lorem
% rhoncus. Maecenas eu arcu ac neque placerat aliquam. Nunc pulvinar
% massa et mattis lacinia.

\end{document}